\documentclass[a4paper]{jpconf}	

\usepackage{braket}
\usepackage{bm}
\usepackage{mathtools,amsmath, amsfonts,amssymb}
\usepackage{mathrsfs}
  \DeclareMathAlphabet{\mathpzc}{OT1}{pzc}{m}{it}
  
\usepackage{textpos}
\newcommand{\kket}[1]{| #1 \rangle\!\rangle}
\newcommand{\bbra}[1]{\langle\!\langle #1 |}
\newcommand{\bbrakket}[1]{\langle\!\langle #1 \rangle\!\rangle}

\newcommand{\phit}{\widetilde{\phi}}
\newcommand{\deltat}{\widetilde{\delta}}
\newcommand{\jt}{\widetilde{j}}
\newcommand{\Deltat}{\widetilde{\Delta}}
\renewcommand{\NC}{\mathrm{NC}}
\newcommand{\CC}{\mathrm{CC}}
\renewcommand{\Im}{\mathrm{Im}}
\renewcommand{\Re}{\mathrm{Re}}
\newcommand{\D}{\mathrm{d}}

\begin{document}

\title{Negative-frequency modes in quantum field theory}
\author{Robert Dickinson$^1$, Jeff Forshaw$^1$ and \underline{Peter Millington}$^2$}
\address{$^1$Consortium for Fundamental Physics,
  School of Physics and Astronomy,
  University of Manchester,
  Manchester M13 9PL,
  United Kingdom}
\address{$^2$Physik Department T70, James-Franck-Stra\ss e,\\
Technische Universit\"{a}t M\"{u}nchen, 85748 Garching, Germany}
\ead{robert.dickinson-2@manchester.ac.uk, jeff.forshaw@manchester.ac.uk, p.w.millington@tum.de}

\begin{textblock}{8}(8.25,-9.7)
\begin{flushright}
\begin{footnotesize}
MAN/HEP/2015/03, TUM-HEP-983-15 \\
March 2015
\end{footnotesize}
\end{flushright}
\end{textblock}

\begin{abstract}
We consider a departure from standard quantum field theory, constructed so as to permit momentum eigenstates of both positive and negative energy. The resulting theory is intriguing because it brings about the cancellation of leading ultra-violet divergences and the absence of a zero-point energy. The theory gives rise to tree-level source-to-source transition amplitudes that are manifestly causal and consistent with standard S-matrix elements. It also leads to the usual result for the oblique corrections to the standard electroweak theory. Remarkably, the latter agreement relies on the breakdown of naive perturbation theory due to resonance effects. It remains to be shown that there are no problems with perturbative unitarity.
\end{abstract}

\section{Introduction}

In 1932, Fermi~\cite{Fermi} calculated the probability for an atom to be excited from its ground state by the absorption of a photon emitted from a second excited atom at a distance $r$. Fermi showed that the minimum time for such a process to occur was $r/c$ in accordance with Einstein causality. However, his result was obtained by replacing an integral over only positive frequencies by an integral over both positive and negative frequencies~\cite{Shirokov, Hegerfeldt:1993qe}. Asymmetry between positive and negative frequencies is a feature of the Feynman propagator, which describes the propagation of positive-frequency modes forwards in time and negative-frequency modes backwards in time. Because of this, it is non-vanishing for spacelike separations, which makes causality at the level of the particle dynamics hard to quantify.

In this note, we will take seriously the potential role of negative-frequency modes propagating forwards in time by departing from standard quantum field theory and allowing for negative-energy states. As a result, we will obtain manifestly-causal tree-level source-to-source transition amplitudes, which are in agreement with the usual S-matrix results. We will also remark on two intriguing features of this construction: specifically, the cancellation of leading ultra-violet (UV) divergences and the absence of a zero-point energy. However, we will note a potentially-serious issue with perturbative unitarity. Even so, as a non-trivial example at loop-level, we will recover the standard result~\cite{Kennedy:1988sn,Peskin:1991sw} for the electroweak oblique corrections of the Standard Model, which relies on the breakdown of naive perturbation theory as a result of resonance effects.


\section{Negative-energy states}

We consider a real scalar field of mass $m$. With the aim of accommodating both positive- and negative-energy states, we introduce two pairs of interaction-picture creation and annihilation operators, satisfying the algebra
\begin{equation}
\label{ops}
\big[a^{\pm}_{\mathbf{p}}(t),\ a^{\pm\dag}_{\mathbf{q}}(t)\big]\ =\ \pm\:2E_{\mathbf{p}}\,\delta^{(3)}_{\mathbf{p},\,\mathbf{q}}\;,\\
\end{equation}
with all other commutators vanishing. Here, the on-shell energy $E_{\mathbf{p}}=(\mathbf{p}^2+m^2)^{1/2}>0$ and we use the shorthand notation $\delta^{(3)}_{\mathbf{p},\,\mathbf{q}}\equiv(2\pi)^3\delta^{(3)}(\mathbf{p}-\mathbf{q})$ for the Dirac delta function.

We wish to interpret both $a^{+\dag}_{\mathbf{p}}$ and $a^{-\dag}_{\mathbf{p}}$ as \emph{creation} operators, with $a^{+}_{\mathbf{p}}$ and $a^{-}_{\mathbf{p}}$ annihilating a vacuum state $\kket{0}$. Thus, the imposition of the algebra in~\eqref{ops} requires us to introduce momentum eigenstates with the following orthonormality:
\begin{subequations}
\label{states}	
\begin{align}
\bbrakket{\mathbf{p}^{\pm}|\mathbf{q}^{\pm}}\ &=\ \pm\:2E_{\mathbf{p}}\,\delta^{(3)}_{\mathbf{p},\,\mathbf{q}}\;,\\
\bbrakket{\mathbf{p}^{\pm}|\mathbf{q}^{\mp}}\ &=\ 0\;.
\end{align}
\end{subequations}
Clearly, if we were to interpret the states in~\eqref{states} as vectors in the standard Hilbert space, those of $-$ type would, having negative norm, give rise to negative probabilities. It is for this reason that we have introduced the double bra-ket notation, which we will describe in Section~\ref{sec:double}.

Combining $+$ and $-$ type contributions, the free Lagrangian density is
\begin{equation}
\label{lag2}
\mathcal{L}^{(0)}_x\ =\ \frac{1}{2}\,\delta_{ab}\big(\partial_{\mu}\phi^{a}_x\partial^{\mu}\phi^{b}_{x}\:-\:m^2\phi^a_x\phi^b_x\big)\;,
\end{equation}
where $\delta_{ab}$ is the Kronecker delta and the indices $a,\:b\in\{1,\:2\}\equiv\{+,\:-\}$ run over the $+$ and $-$ type field operators, given by
\begin{equation}
\label{fields}
\phi^{\pm}_{x}\ \equiv\ \phi^{\pm}(x) \ =\ \pm\:\int_{\mathbf{p}}\frac{1}{2E_{\mathbf{p}}}\;\Big(a^{\pm}_{\mathbf{p}}(0)e^{\mp iE_{\mathbf{p}}x^0}e^{\pm i\mathbf{p}\cdot\mathbf{x}}\:+\:\mathrm{H.c.}\Big)\;,
\end{equation}
where $\mathrm{H.c.}$ indicates the Hermitian-conjugate term and the overall minus sign of $\phi^-_x$ has been chosen by convention. Throughout, we use the shorthand notations
\begin{equation}
\int_p\ \equiv\ \int\!\frac{\D^4 p}{(2\pi)^4}\;,\qquad \int_{\mathbf{p}}\ \equiv\ \int\!\frac{\D^3\mathbf{p}}{(2\pi)^3}\;,\qquad \int_x\ \equiv\ \int\!\D^4x\;,\qquad  \int_{\mathbf{x}}\ \equiv\ \int\D^3\mathbf{x}~.
\end{equation}
The Lagrangian in~\eqref{lag2} is chosen to be symmetric under the interchange of $+$ and $-$ type fields to ensure that the $-$ type momentum states are eigenstates of the free Hamiltonian,
\begin{equation}
\label{H0}
H^{(0)}\ = \ \frac{1}{2}\int_{\mathbf{x}}\;\delta_{ab}\big(\partial_0\phi_x^a\partial_0\phi_x^b
\:+\:\bm{\nabla}\phi^a_x\cdot\bm{\nabla}\phi^b_x\:+\:m^2\phi_x^a\phi_x^b\big)\;,
\end{equation}
with negative eigenvalues: 
\begin{equation}
H^{(0)}\kket{\mathbf{p}^{\pm}}\ =\ \pm\,E_{\mathbf{p}}\kket{\mathbf{p}^{\pm}}\;.
\end{equation}
Substituting the field operators from~\eqref{fields} into~\eqref{H0} and using the algebra from~\eqref{ops}, the free Hamiltonian may be written in the form
\begin{equation}
H^{(0)}\ =\ \frac{1}{2}\int_{\mathbf{p}}\;\Big(a^{+\dag}_{\mathbf{p}}(0)a^{+}_{\mathbf{p}}(0)\:+\:a^{-\dag}_{\mathbf{p}}(0)a^{-}_{\mathbf{p}}(0)\Big)\;.
\end{equation}
We notice that the zero-point contribution has cancelled by virtue of the relative sign between the $+$ and $-$ type commutation relations in~\eqref{ops} (see also~\cite{Elze:2005kt}) and without having imposed normal ordering. Given the democracy imposed between positive- and negative-energy states, we anticipate this cancellation of pure vacuum contributions to be a generic feature.

We now consider the propagators and Wick contractions of the $+$ and $-$ type fields. From~\eqref{ops} and~\eqref{fields}, we obtain the canonical commutation relations
\begin{equation}
\label{comms}
\big[\phi^{\pm}_{x},\ \phi^{\pm}_{y}\big]\ =\ \Delta^{(0)}_{xy}\;,
\end{equation}
where $\Delta^{(0)}_{xy}$ is the free Pauli-Jordan function
\begin{equation}
\Delta^{(0)}_{xy}\ =\ \int_{\mathbf{p}}\frac{1}{2E_{\mathbf{p}}}\;\Big(e^{-iE_{\mathbf{p}}x^0}e^{+i\mathbf{p}\cdot\mathbf{x}}\:-\:e^{+iE_{\mathbf{p}}x^0}e^{-i\mathbf{p}\cdot\mathbf{x}}\Big)\;,
\end{equation}
which vanishes for space-like separations $(x-y)^2<0$. In what follows, we will indicate free propagators by a superscript $(0)$. Notice that the commutation relation~\eqref{comms} is symmetric under interchange of $+$ and $-$ type field operators in correspondence with the symmetry of the Pauli-Jordan function under the transformation $E_{\mathbf{p}}\to-\,E_{\mathbf{p}}$. In addition, we have the following time-ordered propagators:
\begin{equation}
\label{FDprops}
\bbrakket{\mathrm{T}[\phi^{\pm}_{x}\phi^{\pm}_{y}]}\ =\ \begin{cases} +\:\Delta^{\mathrm{F}(0)}_{xy}\ \equiv\ \Delta^{+(0)}_{xy}\\-\:\Delta^{\mathrm{D}(0)}_{xy}\ \equiv\ \Delta^{-(0)}_{xy}\end{cases}\!\!\!\!\!\!,
\end{equation}
where $\bbrakket{\bullet}\:\equiv\:\bbrakket{0|\bullet|0}$ and
\begin{equation}
\Delta^{\mathrm{F}[\mathrm{D}](0)}_{xy}\ =\ [-]\int_p\frac{ie^{-ip\cdot(x-y)}}{p^2-m^2+[-]i\epsilon}\;,
\end{equation}
is the Feynman [Dyson] propagator, with $\epsilon\to 0^+$. The Wightman propagators are given by
\begin{equation}
\Delta^{\gtrless(0)}_{xy}\ =\ \pm\:\bbrakket{\phi^{\pm}_{x}\phi^{\pm}_{y}}\ =\ \mp\:\bbrakket{\phi^{\mp}_{y}\phi^{\mp}_{x}}\;,
\end{equation}
and the Hadamard propagator by
\begin{equation}
\Delta^{1(0)}_{xy}\ =\ \pm\:\bbrakket{\{\phi^{\pm}_{x},\ \phi^{\pm}_{y}\}}\ =\ \Delta^{>(0)}_{xy}\:+\:\Delta^{<(0)}_{xy}\;,
\end{equation}
where $\{\,,\}$ denotes the anti-commutator. The retarded and advanced propagators are
\begin{equation}
\Delta^{\mathrm{R}(0)}_{xy}\ =\ \theta_{xy}\,\Delta^{(0)}_{xy}\ =\ \Delta^{\mathrm{A}(0)}_{yx}\;,
\end{equation}
where $\theta_{xy}\equiv\theta(x^0-y^0)$ is the unit step function. We also define the principal-part propagator:
\begin{equation}
\label{ppprop}
\Delta^{\mathcal{P}(0)}_{xy}\ =\ \frac{1}{2}\big(\Delta^{\mathrm{R}(0)}_{xy}\:+\:\Delta^{\mathrm{A}(0)}_{xy}\big)\;.
\end{equation}
Finally, we note the field-particle duality relations
\begin{subequations}
\label{fieldpart}
\begin{align}
\bbrakket{\phi^{\pm}_xa^{\pm\dag}_{\mathbf{p}}(t)}\ &=\ e^{\mp i E_{\mathbf{p}}(x^0-t)}e^{\pm i\mathbf{p}\cdot \mathbf{x}}\;,\\
\bbrakket{a^{\pm}_{\mathbf{p}}(t)\phi^{\pm}_x}\ &=\ e^{\pm i E_{\mathbf{p}}(x^0-t)}e^{\mp i\mathbf{p}\cdot \mathbf{x}}\;.
\end{align}
\end{subequations}

By inspection, we see that the various propagators of the $+$ and $-$ type field operators are related by the transformation $\epsilon\to -\,\epsilon$, i.e.~by the interchange of the Feynman and Dyson pole prescriptions. We recall that the Feynman prescription places the poles of the propagator in the second and fourth quadrants of the complex plane, describing the propagation of positive-frequency modes forwards in time and negative-frequency modes backwards in time. Conversely, the Dyson prescription places the poles in the first and third quadrants, describing the propagation of negative-frequency modes forwards in time and positive-frequency modes backwards in time. Thus, this construction is manifestly symmetric under the discrete symmetry transformation $E_{\mathbf{p}}\to -\,E_{\mathbf{p}}$. We note that, using the limit representation
\begin{equation}
\delta(x)\ =\ \lim_{\epsilon \to 0^+} \frac{1}{\pi}\,\frac{\epsilon}{x^2+\epsilon^2}
\end{equation}
of the Dirac delta function, the transformation of the Wightman propagators under the interchange of $+$ and $-$ field operators also follows from the transformation $\epsilon\to -\,\epsilon$. The retarded, advanced and principal-part propagators, being related to the Pauli-Jordan function, are symmetric under this transformation.

\section{Doubled Hilbert space}
\label{sec:double}

Having imposed democracy between positive- and negative-frequency modes, we now consider a mathematical structure that permits such a construction.

Since we have required that the $+$ and $-$ type operators commute and that the sets of $+$ and $-$ type momentum eigenstates are each complete, it follows that the $+$ and $-$ operators must be confined to separate Hilbert spaces. Hence, we consider a product space $\widehat{\mathscr{H}}=\mathscr{H}\times\mathscr{H}^{T}$, with the $+$ type operators corresponding to the trivial embedding of the usual scalar creation and annihilation operators
\begin{equation}
a^{+(\dag)}_{\mathbf{p}}(t)\ \equiv\ a_{\mathbf{p}}^{(\dag)}(t)\:\otimes\:\mathbb{I}\;.
\end{equation}
This construction, although having a very different interpretation, is similar to the doubled Hilbert space of thermo field dynamics (TFD)~\cite{Takahasi:1974zn, Arimitsu:1985xm}.

In order to accommodate the $-$ type states, we supplement the field of complex numbers $\mathbb{C}$, having the usual imaginary unit $i\equiv i_1$, with an additional imaginary unit $i_2$. Defining the involutions $\ast$ and $\star$ to be complex conjugation with respect to $i_1$ and $i_2$, respectively, we have the following properties:
\begin{subequations}
\begin{gather}
i_1^*\ =\ -\,i_1\;,\qquad i_1^{\star}\ =\ +\,i_1\;, \qquad i_1^2\ =\ -\,1\;,\\
i_2^*\ =\ +\,i_2\;,\qquad i_2^{\star}\ =\ -\,i_2\;, \qquad i_2^2\ =\ -\,1\;.
\end{gather}
\end{subequations}
Together with the real unit $1$ and the product $i_1i_2=i_2i_1$, we have a four-dimensional basis $\{1,\:i_1,\:i_2,\:i_1i_2\}$ that spans the ring of bicomplex numbers (or tessarines) $\mathbb{T}$, forming a commutative and associative algebra over the real numbers $\mathbb{R}$, see e.g.~\cite{bicomplex}. Given two hypercomplex vectors $\mathbf{u}$ and $\mathbf{v}$, we may construct three distinct scalar products:
\begin{equation}
\mathbf{u}^{\mathsf{T}*}\mathbf{v}\ =\ a \:+\:i_2\,b\;,\qquad
\mathbf{u}^{\mathsf{T}\star}\mathbf{v}\ =\ c \:+\:i_1\,d\;,\qquad
\mathbf{u}^{\mathsf{T}*\star}\mathbf{v}\ =\ e\ \geq \ 0\;,
\end{equation}
where $a,\:b,\:c,\:d,\:e\in\mathbb{R}$. For $\mathbf{v}=\mathbf{u}$, these scalar products correspond to the modulus with respect to $i_1$, the modulus with respect to $i_2$ and the $\mathbb{T}$-norm, respectively. We choose to define the inner product of states
\begin{equation}
\bbrakket{\mathbf{p}^{\pm}|\mathbf{q}^{\pm}} \equiv (\kket{\mathbf{p}^{\pm}})^{\dag}\kket{\mathbf{q}^{\pm}}\;,
\end{equation}
with the Hermitian conjugate ($\dag$) being the transpose of the complex conjugate with respect to $i_1$ only.\footnote{Rather than the usual $C*$-algebra over the field of complex numbers, our operator algebra corresponds to a $*$-algebra over the ring of bicomplex numbers, with the antiautomorphic involution provided by the Hermitian conjugate with respect to $i_1$. It would be of interest to compare this construction with $PT$-symmetric theories, see e.g.~\cite{Bender:2005tb}, in which a non-positive inner product can be defined with respect to $PT$ transformations.}

We may now define the $-$ type creation and annihilation operators by the following discrete transformation:
\begin{equation}
\label{aminusdef}
a^-_{\mathbf{p}}(t)\ \equiv\ -\,i_2\,(a^{+}_{\mathbf{p}}(t))^{T_Q}\ =\ -\,i_2\,\widehat{\mathcal{U}}_T\,Q\,a^{+}_{\mathbf{p}}(t)\,Q^{\mathsf{T}}\,\widehat{\mathcal{U}}_T^{-1}\ =\ -\,i_2\,\mathbb{I}\:\otimes\: a_{\mathbf{p}}^*(t) \;,
\end{equation}
where $Q$ is a permutation matrix that reorders the Kronecker product and $\widehat{\mathcal{U}}_T=\mathcal{U}_T\otimes\mathcal{U}_T$, with $\mathcal{U}_T$ akin to the usual time-reversal operator in Fock space, but defined instead to be anti-linear with respect to both $i_1$ and $i_2$, i.e.~for $\lambda\in\mathbb{T}$, $\mathcal{U}_T\lambda\,a_{\mathbf{p}}\,\mathcal{U}_T^{-1} = \lambda^{*\star}\,a^*_{\mathbf{p}}$. Thus, we may quickly verify that it is involutory, i.e.~$[(a_{\mathbf{p}}^\pm)^{T_Q}]^{T_Q}=a_{\mathbf{p}}^{\pm}$. Choosing the overall sign by convention, the $+$ and $-$ type field operators are related via
\begin{equation}
\label{phiminusdef}
\phi^-_x\ \equiv\ i_2\,(\phi^+_x)^{T_Q}\ =\ i_2\,\widehat{\mathcal{U}}_T\,Q\, \phi^{+}_{x}\,Q^{\mathsf{T}}\,\widehat{\mathcal{U}}_T^{-1}\;.
\end{equation}
We may then show that the free Lagrangian in~\eqref{lag2} is anti-symmetric under $T_Q$, i.e.
\begin{equation}
\label{trans}
(\mathcal{L}^{(0)}_{x})^{T_Q} \ =\ -\,\mathcal{L}^{(0)}_{x}\;.
\end{equation}
As we will see, the role of the factors of $i_2$ in this construction is solely to produce systematically the relative signs between $+$ and $-$ contributions, consistent with the interchange $E_{\mathbf{p}}\to-\,E_{\mathbf{p}}$.

We introduce the vacuum state $\kket{0} \equiv \ket{0}\otimes\ket{0}$ of $\widehat{\mathscr{H}}$,
satisfying $\widehat{\mathcal{U}}_T\,Q\,\kket{0} = \kket{0}$.
We then have the following actions for the creation and annihilation operators:
\begin{equation}
a^{\pm\dag}_{\mathbf{p}}(t)\kket{0}\ =\ \kket{\mathbf{p}^{\pm}(t)}\;,\qquad 
a^{\pm}_{\mathbf{p}}(t)\kket{\mathbf{q}^{\pm}(t)} \ =\ \pm\,2E_{\mathbf{p}}\,\delta^{(3)}_{\mathbf{p},\,\mathbf{q}}\kket{0}\;,
\end{equation}
where
\begin{subequations}
\begin{align}
\kket{\mathbf{p}^+(t)} &= a_{\mathbf{p}}^{\dag}(t)\ket{0}\:\otimes\:\ket{0}=\ket{\mathbf{p}(t)}\:\otimes\:\ket{0}\;,\\
\kket{\mathbf{p}^-(t)} &= -\,i_2\ket{0}\:\otimes\: \mathcal{U}_T\,a^{\dag}_{\mathbf{p}}(t)\,\mathcal{U}^{-1}_T\ket{0} = -\,i_2\ket{0}\:\otimes\:\ket{\mathbf{p}(t)}^*\;,
\end{align}
\end{subequations}
consistent with~\eqref{ops} and~\eqref{states}.
Finally, the completeness relations of the $+$ and $-$ Fock bases may be written as
\begin{equation}
\mathbb{I}^{\pm}\ =\ \kket{0}\bbra{0}\:\pm\:\int_{\mathbf{p}_1}\!\frac{1}{2E_{\mathbf{p}_1}}\,\kket{\mathbf{p}^{\pm}_1}\bbra{\mathbf{p}^{\pm}_1}\:+\:\frac{1}{2!}\!\int_{\mathbf{p}_1}\!\frac{1}{2E_{\mathbf{p}_1}}\int_{\mathbf{p}_2}\!\frac{1}{2E_{\mathbf{p}_2}}\,\kket{\mathbf{p}^{\pm}_1,\mathbf{p}^{\pm}_2}\bbra{\mathbf{p}^{\pm}_1,\mathbf{p}^{\pm}_2}\pm\cdots\;,
\end{equation}
in which the $n$-particle state is $\kket{\mathbf{p}_1^{\pm},\mathbf{p}_2^{\pm},\dots,\mathbf{p}_n^{\pm}}\equiv\kket{\mathbf{p}_1^{\pm}}\otimes
\kket{\mathbf{p}_2^{\pm}}\otimes\cdots\otimes\kket{\mathbf{p}_n^{\pm}}$.

\section{Matrix elements}
\label{sec:matrix}

Were we to redefine $a^{-(\dag)}_{\mathbf{p}}\to i_2\,a^{-(\dag)}_{\mathbf{p}}$ in the preceding sections, we would immediately recover the algebra of TFD~\cite{Takahasi:1974zn, Arimitsu:1985xm}, with the transformation $T_Q$ corresponding to the operation of ``tilde-conjugation''. The free Hamiltonian in~\eqref{H0} would then be the Liouvillian operator, whose role is to generate time-translations of ``thermal'' states. Instead, we wish to consider interactions that preserve the democracy of the Hamiltonian in the $+$ and $-$ type field operators. For this reason, we \emph{choose} to consider the following interaction Hamiltonian density:
\begin{equation}
\label{intham}
\widehat{\mathcal{H}}_{J,\,x}^{\mathrm{int}}\ =\ \frac{g}{3!}\big[(\phi^+_x)^3\:+\:(\phi^-_x)^3\big]\:-\:J_x\big[\phi^+_x\:+\:\phi^-_x\big]\;,
\end{equation}
where $J_x$ is a real-valued classical source and $g\in\mathbb{R}$ is a coupling of mass dimension one. We have not allowed for terms that directly mix $+$ and $-$ contributions, since these would potentially lead to an ill-defined perturbation series, containing products of Dirac delta functions with identical arguments, see e.g.~\cite{Millington:2012pf}.

We might be concerned that the interaction Hamiltonian density in~\eqref{intham} is complex-valued with respect to both $i_1$ and $i_2$. However, by Wick's theorem, any non-vanishing contractions of $-$ type field operators will be built from the expectation values of bi-linears. Since $\phi^-_x$ is purely imaginary with respect to $i_2$, i.e.~$\phi^{-\star}_x=-\,\phi^-_x$, all such correlation functions will contain only even powers of $i_2$. Thus, all correlation functions and the corresponding matrix elements will be complex-valued with respect to $i_1$ only, giving real-valued and positive-semi-definite probabilities.

Since the classical source $J_x$ couples only to the sum $\phi^+_x+\phi^-_x$, we consider the following change of basis:
\begin{equation}
\label{tildefield}
\phit^{\pm}_{x}\ \equiv\ \phi^{+}_{x}\:\pm\:\phi^{-}_{x}\;,
\end{equation}
which resembles the orthogonal rotation to the so-called Keldysh (or physical) basis of thermal field theory, see Section~\ref{sec:path}, used in~\cite{Dickinson:2013lsa} to make apparent the causal structure of source-to-sink amplitudes. In this basis, we may write the Hamiltonian density in the form
\begin{equation}
\label{physH}
\widehat{\mathcal{H}}^{\mathrm{int}}_{J,\,x}\ =\ \frac{g}{4!}\Big[\big(\phit^+_x\big)^3\:+\:3\,\phit^+_x\big(\phit^-_x\big)^2\Big]\:-\:J_x\,\phit^+_x\;,
\end{equation}
with the classical source $J_x$ now coupling only to the field $\phit^+_x$. Note the factor of $4!$ for the coupling, which results from the fact that we have not chosen $\phit^{\pm}_x$ to be canonically normalized.

The two-point correlation functions split into two sets: those containing only one of $\phit^+_x$ and $\phit^-_x$, which are entirely \emph{causal}, and those containing both $\phit^+_x$ and $\phit^-_x$, which are entirely \emph{acausal}:
\begin{subequations}
\begin{align}
\Delta^{(0)}_{xy}\ &=\ \frac{1}{2}\,[\phit^{\pm}_{x},\ \phit^{\pm}_{y}]\ =\ \varepsilon_{xy}\bbrakket{\mathrm{T}[\phit^{\pm}_{x}\phit^{\pm}_{y}]}\ = \ \bbrakket{\phit^{\pm}_{x}\phit^{\pm}_{y}}\ =\ -\:\bbrakket{\phit^{\pm}_{y}\phit^{\pm}_{x}}\;,\\
\Delta^{1(0)}_{xy}\ &=\ \frac{1}{2}\,\bbrakket{\{\phit^{\pm}_{x},\ \phit^{\mp}_{y}\}}\ =\ \bbrakket{\mathrm{T}[\phit^{\pm}_{x}\phit^{\mp}_{y}]}\ = \ \bbrakket{\phit^{\pm}_{x}\phit^{\mp}_{y}}\ =\ +\:\bbrakket{\phit^{\pm}_{y}\phit^{\mp}_{x}}\;,
\end{align}
\end{subequations}
where $\varepsilon_{xy}\:=\:\theta_{xy}\:-\:\theta_{yx}$ is the signum function. For later convenience, we introduce the following notation:
\begin{subequations}
\begin{align}
\Deltat^{+(0)}_{xy} &\equiv \bbrakket{\mathrm{T}[\phit^{\pm}_{x}\phit^{\pm}_{y}]} \ =\  2\,\Delta^{\mathcal{P}(0)}_{xy}\ =\  2i\mathcal{P}\!\int_p\frac{e^{-ip\cdot(x-y)}}{p^2-m^2}\;,\\
\Deltat^{-(0)}_{xy} &\equiv \bbrakket{\mathrm{T}[\phit^{\pm}_{x}\phit^{\mp}_{y}]} \ = \ \Delta_{xy}^{1(0)} \ =\  \int_p e^{-ip\cdot(x-y)}\;2\pi\delta(p^2-m^2)\;,
\end{align}
\end{subequations}
where $\mathcal{P}$ denotes the Cauchy principal value.

In order to define asymptotic states, we consider the equations of motion of the (Heisenberg-picture) \emph{in} field operators
\begin{equation}
\label{Jfield}
Z_{\phi}^{1/2}\big(\Box_x+m^2\big)\phi^{\pm}_{\mathrm{in},x}\ =\ J_x\;.
\end{equation}
Since the free field operator $\phi^-_x$ is complex-valued with respect to $i_2$, using~\eqref{phiminusdef} and assuming $J_x^{T_Q}=J_x$, the \emph{in} field operators are necessarily defined via
\begin{equation}
\label{infield}
\phi_{\mathrm{in},x}^{+}\ \approx\ Z_{\phi}^{-1/2}\phi^+_x\;,\qquad \phi_{\mathrm{in},x}^{-}\ \approx\ -\,i_2Z_{\phi}^{-1/2}\phi^-_x\;,
\end{equation}
with
\begin{equation}
a^{+(\dag)}_{\mathrm{in},\mathbf{p}}\ \approx\ Z_{\phi}^{-1/2}a_{\mathbf{p}}^{+(\dag)}(0)\;,\qquad a^{-(\dag)}_{\mathrm{in},\mathbf{p}}\ \approx\ +\,i_2Z_{\phi}^{-1/2}a_{\mathbf{p}}^{-(\dag)}(0)\;,
\end{equation}
by virtue of~\eqref{aminusdef}. For generating the external states, we consider the coherent state
\begin{equation}
\kket{\alpha(t)}\ =\ \exp\bigg[-\frac{1}{2}\!\int_{\mathbf{k}}\frac{1}{2E_{\mathbf{k}}}\,\alpha_{\mathbf{k}}^*\alpha_{\mathbf{k}}\bigg]\exp\!\bigg[\int_{\mathbf{k}}\frac{1}{2E_{\mathbf{k}}}\,\alpha_{\mathbf{k}}\Big(a^{+\dag}_{\mathbf{k}}(t)+ a^{-\dag}_{\mathbf{k}}(t)\Big)\bigg]\kket{0}\;,
\end{equation}
where $\alpha_{\mathbf{k}}$ is a complex-valued function with respect to $i_1\equiv i$. The combination of operators in the exponent is democratic in the $+$ and $-$ contributions in accordance with the coupling of the classical source $J_x$. Hence, by repeated functional differentiation with respect to $\alpha_{\mathbf{k}}$, we can generate the $n$-particle state $\kket{\widetilde{\mathbf{p}}_{1,\dots,\,n}}=\kket{\widetilde{\mathbf{p}}_{1}}\otimes
\kket{\widetilde{\mathbf{p}}_{2}}\otimes \cdots\otimes \kket{\widetilde{\mathbf{p}}_{n}}$, where
\begin{equation}
\label{tildestate}
\kket{\widetilde{\mathbf{p}}(t)}\ \equiv\ \frac{\delta}{\delta\alpha_{\mathbf{p}}}\,\exp\bigg[\frac{1}{2}\!\int_{\mathbf{k}}\frac{1}{2E_{\mathbf{k}}}\,\alpha_{\mathbf{k}}^*\alpha_{\mathbf{k}}\bigg]\kket{\alpha(t)}\bigg|_{\alpha_{\mathbf{k}}\, =\, 0}\ = \ \kket{\mathbf{p}^+(t)}\:+\:\kket{\mathbf{p}^-(t)}\;.
\end{equation}
The corresponding asymptotic state $\kket{\widetilde{\mathbf{p}}_{\mathrm{in}}}
=\kket{\mathbf{p}^+_{\mathrm{in}}}
-i_2\,\kket{\mathbf{p}^-_{\mathrm{in}}}$ has the following contractions
\begin{equation}
\bbrakket{0|\phit^{\pm}_x|\widetilde{\mathbf{p}}_{\mathrm{in}}}\ = \
\begin{cases} 2Z_{\phi}^{-1/2}\cos(\mathbf{p}\cdot\mathbf{x}-E_{\mathbf{p}}x_0)\\2iZ_{\phi}^{-1/2}\sin(\mathbf{p}\cdot\mathbf{x}-E_{\mathbf{p}}x_0)\end{cases}\!\!\!\!\!\!\!\;.
\end{equation}

The scattering-matrix operator is
\begin{equation}
\widehat{\mathcal{S}}\ =\ \mathrm{T}\exp\bigg[-i\!\int_{-\infty}^{+\infty}\D^4 x\;\widehat{\mathcal{H}}^{\mathrm{int}}_x\bigg]\;,
\end{equation}
where $\widehat{\mathcal{H}}^{\mathrm{int}}_x$ is the interaction Hamiltonian density in~\eqref{intham} with the source $J_x$ set to zero. Defining the transition operator $i\widehat{\mathcal{T}}=\widehat{\mathcal{S}}-\mathbb{I}$ in the usual way, the $m\to n$ particle matrix element is
\begin{equation}
\label{M}
i\mathcal{M}_{m\to n}\ =\ \bbrakket{\widetilde{\mathbf{q}}_{\mathrm{in},\,1,\,\dots,\, n}|\:i\widehat{\mathcal{T}}\:|\widetilde{\mathbf{p}}_{\mathrm{in},\,1,\,\dots,\,m}}\;.
\end{equation}
Since there is no direct mixing between $+$ and $-$ contributions, it is easiest to perform the calculation in the non-tilde basis. In this case, \eqref{M} corresponds to the following rule:
\vspace*{0.5cm} \\
\noindent\framebox[\textwidth][t]{\hspace{0.75em}\begin{minipage}[]{\textwidth - 2.2em}\vspace{0.75em}
Calculate the standard matrix element $i\mathcal{M}^+$, containing only free Feynman propagators $\Delta_{\mathrm{F}}^{(0)}$, and add to it the matrix element $i\mathcal{M}^-$, with all $\Delta_{\mathrm{F}}^{(0)}$ replaced by $-\,\Delta_{\mathrm{D}}^{(0)}$, i.e.
\begin{equation}
i\mathcal{M}\ \equiv\ i\mathcal{M}^+\:+\:i\mathcal{M}^-\;.
\end{equation}
Equivalently, sum a given matrix element over both choices of pole prescription at the amplitude level, with $i\mathcal{M}^+$ corresponding to the Feynman prescription $m^2\to m^2-i\epsilon$ and $i\mathcal{M}^-$ corresponding to the Dyson prescription $m^2\to m^2+i\epsilon$.\vspace{0.75em}\end{minipage}}
\vspace*{0.5cm} \\
The immediate implication of this rule is that, whilst all tree-level matrix elements ($i\mathcal{M}$) are purely imaginary, all one-loop matrix elements are purely real (and so on). \emph{Prima facie}, it would therefore appear that there are serious problems with the viability (both phenomenological and theoretical) of the whole venture. For example,
there can be no interference between tree-level and one-loop diagrams, which presents a problem for perturbative unitarity, e.g.~the Bloch-Nordsieck~\cite{Bloch:1937pw} cancellation of infra-red divergences in gauge theories. Even so, as we shall see in Section~\ref{sec:EW}, there is a breakdown of naive perturbation theory, which means that we are able to recover the ``one-loop'' oblique corrections in the Standard Model. This agreement seems quite remarkable to us and, for that reason, we shall press on.
 
By way of illustration,\footnote{The $1\to2$ process is, of course, kinematically disallowed.} the rule leads to the following matrix element for the $1\to 2$ process:
\begin{align}
i\mathcal{M}_{1\to 2}\ &=\ \frac{1}{3!}\,(-ig)\!\int_x\; \bbrakket{\widetilde{\mathbf{q}}_{\mathrm{in},\,1,\,2}|\big[(\phi^+_x)^3+(\phi^-_x)^3\big]|\widetilde{\mathbf{p}}_{\mathrm{in}}} \nonumber \\
&=\ 2Z_{\phi}^{-3/2}(-ig)\,\delta^{(4)}_{q_1+q_2,\,p}
\label{1to2tree}
\end{align}
and the $2\to 2$ matrix element is
\begin{equation}
\label{2to2tree}
i\mathcal{M}_{2\to2}\ =\ 2Z_{\phi}^{-2}(-ig)^2\,\delta^{(4)}_{q_1+q_2,\,p_1+p_2}\sum_{k^2\,=\,s, t, u}\Delta^{\mathcal{P}(0)}_{k^2}\;,
\end{equation}
where $s$, $t$ and $u$ are the Mandelstam variables. Since the pole prescription of the intermediate propagator is irrelevant at tree-level, \eqref{1to2tree} and~\eqref{2to2tree} appear to be consistent with the usual S-matrix results, except that they disagree by an overall factor of $2$. The latter is, however, not the case.

We recall that the classical source $J_x$ couples to $\phit^+_x$. Since we will later use this classical source to perform the LSZ reduction~\cite{Lehmann:1954rq} of $n$-point Green's functions, we should define the wavefunction renormalization $Z_{\phi}$ with respect to the field $\phit^+_x$. At leading order, $Z_{\phi}$ is obtained by expanding
\begin{equation}
\Deltat^{+(0)}_{k^2}\ =\ \lim_{\epsilon\to 0^+}\int_k\frac{2i\big(k^2-m^2\big)e^{-ik\cdot(x-y)}}{\big(k^2-m^2\big)^2\:+\:\epsilon^2}
\end{equation}
about the root $\overline{m}^2$ of $\big[(k^2-m^2)^2+\epsilon^2\big]_{k^2\,=\,\overline{m}^2} = 0$. Trivially, $\overline{m}^2=m^2$ in the limit $\epsilon\to0^+$, and we find $ Z_{\phi}=2$. Since $\overline{m}^2=Z_m^{-1}Z_{\phi}m^2$, it follows that the mass renormalization is $Z_m=2$. In addition, choosing the renormalization condition
\begin{equation}
\frac{\partial^3 V(\phi_x^{\pm})}{\partial(\phit^{+}_x)^3}\ = \  \frac{g}{Z_g}\;,
\end{equation}
where $V(\phi_x^{\pm})$ is the potential, we find $Z_g=4$. The tree-level renormalized coupling is therefore given by $\bar{g}=Z_{g}^{-1}Z_{\phi}^{3/2}g =  g/\sqrt{2}$.

In terms of these renormalized parameters, the $1\to 2$ and $2\to 2$ matrix elements are
\begin{subequations}
\begin{align}
i\mathcal{M}_{1\to 2}\ &=\ (-i\bar{g})\,\delta^{(4)}_{q_1+q_2,\,p}\;,\\
i\mathcal{M}_{2\to 2}\ &=\ (-i\bar{g})^2\,\delta^{(4)}_{q_1+q_2,\, p_1+p_2}\sum_{k^2\,=\, s, t, u}\Delta^{\mathcal{P}(0)}_{k^2}\;,
\end{align}
\end{subequations}
in agreement with the usual results.

For a general tree-level graph with $N$ external legs, $V$ vertices and $M=V-1$ internal legs, $V=N-2$ and each of the $N$ external states contributes $Z_{\phi}^{-1/2}$ to the prefactor of the matrix element, i.e.
\begin{align}
&2Z_{\phi}^{-N/2}(-ig)^V\ = \ (-i\bar{g})^V\;,
\end{align}
in which the overall factor of $2$ is systematically absorbed.

Lastly, in terms of the renormalized parameters and fields, denoted $\phit_{R,x}^{\pm}$, the Lagrangian may be written as
\begin{align}
\mathcal{L}_x\ &=\ \frac{1}{4}\,\delta_{ab}\big(Z_{\phi}\,\partial_{\mu}\phit_{R,x}^{a}\,\partial^{\mu}\phit_{R,x}^{b}-Z_m\overline{m}^2\,\phit^{a}_{R,x}\,\phit^{b}_{R,x}\big)-\frac{1}{4!}\,Z_g\bar{g}\big[(\phit_{R,x}^{+})^3+3\,\phit^+_{R,x}(\phit^-_{R,x})^2\big]+\bar{J}_x\phit_{R,x}^{+}\;,
\end{align}
where $\bar{J}_x \equiv  Z_{\phi}^{1/2}J_x$ is the renormalized source. Inserting the tree-level values, we have
\begin{align}
\mathcal{L}_x\ &=\ \frac{1}{2}\,\delta_{ab}\big(\partial_{\mu}\phit_{R,x}^{a}\,\partial^{\mu}\phit_{R,x}^{b}\:-\:\overline{m}^2\,\phit^{a}_{R,x}\,\phit^{b}_{R,x}\big)\:-\:\frac{1}{3!}\,\bar{g}\big[(\phit_{R,x}^{+})^3+3\,\phit^+_{R,x}(\phit^-_{R,x})^2\big]\:+\:\bar{J}_x\phit_{R,x}^{+}\;,
\end{align}
in which both $\phit^{\pm}_{R,x}$ now appear canonically normalized.

\section{Path-integral representation}
\label{sec:path}

Starting from the vacuum persistence amplitude in the presence of a set of test sources $j^{\pm}_{x}$, $Z[j^{\pm}] = {}_{j^{\pm}}\bbrakket{0|0}_{j^{\pm}}$, we obtain the generating functional
\begin{equation}
Z[j^{\pm}]\ =\ \!\int\![\D\phi^a]\;\exp\bigg[iS[\phi^a]\:+\:i\!\int_x\delta_{ab}\,j_x^a\,\phi_x^b\bigg]\;,
\end{equation}
where the action is given by
\begin{equation}
S[\phi^a]\ =\ \!\int_x\bigg[\frac{1}{2}\,\phi^{a}_x\,\Delta^{(0)-1}_{ab,\,x}\,\phi^{b}_x\:-\:\frac{g}{3!}\,\delta_{abc}\,\phi^{a}_x\,\phi^{b}_x\,\phi^{c}_x\:+\:\delta_{ab}\,J^{a}_x\,\phi^{b}_x\bigg]\;.
\end{equation}
Here, $\Delta^{(0)-1}_{ab,\,x}=-\,(\Box_x\:+\:m^2)\delta_{ab}$ is the Klein-Gordon operator, $J^{a}_{x}\:=\: J_x$ is the classical source and $\delta_{abc}$ is defined to be $1$ when $a\:=\:b\:=\:c$ and $0$ otherwise. This generating functional is similar in form to the path-integral of the Schwinger-Keldysh CTP formalism~\cite{Schwinger:1960qe,Keldysh:1964ud} of thermal field theory, see e.g.~\cite{Millington:2012pf,Berges:2004yj}. However, there is no overall sign between $+$ and $-$ contributions and, more significantly, the matrix propagator $\Delta^{(0)ab}_{xy}$ takes the form
\begin{equation}
\Delta^{(0)ab}_{xy}\ =\ \mathrm{diag}\big(\Delta^{\mathrm{F}(0)}_{xy}\;,\ -\:\Delta^{\mathrm{D}(0)}_{xy}\big)\;,
\end{equation}
with zero off-diagonal elements, rather than Wightman propagators, as is the case in the CTP formalism.

Performing an orthogonal rotation to the physical basis, equivalent to the redefinition in~\eqref{tildefield} of Section~\ref{sec:matrix}, the matrix propagator takes the form
\begin{equation}
\Deltat^{(0)ab}_{xy}\ =\ \begin{bmatrix} \Deltat^{+(0)}_{xy} & \Deltat^{-(0)}_{xy} \\ \Deltat^{-(0)}_{xy} & \Deltat^{+(0)}_{xy}\end{bmatrix}\;.
\end{equation}
Thus, after completing the square in the exponent of the generating functional, we obtain
\begin{equation}
Z[j^{\pm}]\ =\ \exp\!\bigg[\int_xJ_x\,\deltat_x^+\bigg]e^{W[j^{\pm}]}\;,
\end{equation}
where
\begin{align}
W[j^{\pm}]\ =\ \ln\!\Bigg[Z_0[0]\exp\bigg\{\frac{g}{4!}\int_x\Big[\big(\deltat^+_x\big)^3\:+\:3\big(\deltat^-_x\big)^2\deltat^+_x\Big]\!\bigg\}
\exp\bigg\{-\frac{1}{2}\!\int_{xy}\;\jt_x^{a}\,\widetilde{\Delta}^{(0)ab}_{xy}\,\jt_y^{b}\bigg\}\Bigg]
\end{align}
is the generating functional of connected Green's functions, $Z_0[0]$ is the generating functional in the absence of sources and interactions, and $\widetilde{\delta}^{\pm}_x$ indicates functional derivatives with respect to the sources, $\widetilde{j}^{\pm}_x=(j^+_x\pm j^-_x)/2$. The $n$-point connected source-to-source amplitude is
\begin{equation}
\label{GammaJ}
\Gamma_J^{(n)}\ =\ \frac{1}{n!}\Bigg[\prod_{i=1}^{n} \int_{x_i}J_{x_i}\,\deltat_{x_i}^+\Bigg]W[j^{\pm}]\,\Bigg|_{\jt^{\pm}\,=\,0}\;.
\end{equation}

In order to obtain matrix elements, we can effect LSZ reduction~\cite{Lehmann:1954rq} by promoting $J_x$ to an operator in Fock space, before sandwiching~\eqref{GammaJ} between \emph{in} states. Using \eqref{Jfield}, we make the following replacement of $J_x$ in terms of $\phi^+_{\mathrm{in},x}$:
\begin{equation}
\label{Jdef}
J_x\ \to \ Z^{1/2}_{\phi}\phi^+_{\mathrm{in},x}\big(\Box_x\:+\:m^2\big)\;.
\end{equation}
Notice that for plane-wave source-to-source amplitudes, any diagram with $\widetilde{\Delta}^{-(0)}_{xy}\equiv\Delta^{1(0)}_{xy}$ appearing as an external leg will vanish.

In the case of $2\to2$ scattering, only the $(\phit^+_x)^3$ vertex will contribute and the relevant part of the tree-level source-to-source amplitude is 
\begin{equation}
\label{stos}
\Gamma^{(4)}_{J}\ =\ \frac{1}{4}\,(-ig)^2\int_{x_1x_2x_3x_4yz}J_{x_1}J_{x_2}J_{x_3}J_{x_4}\Delta^{\mathcal{P}(0)}_{x_1y}\Delta^{\mathcal{P}(0)}_{x_2y}\Delta^{\mathcal{P}(0)}_{yz}\Delta^{\mathcal{P}(0)}_{zx_3}\Delta^{\mathcal{P}(0)}_{zx_4}\;,
\end{equation}
differing by the expected factor of $2$ from the standard 4-point source-to-source Green's function. Since this source-to-source amplitude is mediated by the principal-part propagator, it is manifestly causal. Using \eqref{Jdef} and sandwiching~\eqref{stos} between two-particle \emph{in} states, we obtain the matrix element
\begin{align}
&i\mathcal{M}_{2\to2}\ =\ \bbrakket{\widetilde{\mathbf{q}}_{\mathrm{in},\,1,\,2}|\Gamma^{(4)}_J|\widetilde{\mathbf{p}}_{\mathrm{in},\,1,\,2}}\  =\ 2Z_{\phi}^{-2}(-ig)^2\delta^{(4)}_{q_1+q_2,\, p_1+p_2}\sum_{k^2\,=\,s,t,u}\Delta^{\mathcal{P}(0)}_{k^2}\;,
\end{align}
in agreement with~\eqref{2to2tree}. It remains to be shown whether this manifestly-causal structure survives at the loop level.

\section{Loop structure}
\label{sec:loops}

We now consider the one-loop correction to the two-point source-to-source amplitude, given by
\begin{equation}
\Gamma_J^{(2)}\ =\ \frac{1}{2}\int_{xy}J_x\,J_y\,\deltat_x^+\,\deltat_y^+\,W[j^{\pm}]\ =\ \frac{1}{2}\int_{xy}J_x\,J_y\Big(\Deltat_{xy}^{+(0)}\:+\:\Deltat_{xy}^{+(1)}\:+\:\cdots\Big)\;.
\end{equation}
Here, $\Deltat_{xy}^{+(1)}$ is the one-loop contribution to the $\Deltat^+_{xy}$ propagator, given by
\begin{equation}
\Deltat^{\pm(1)}_{xy} \ =\  \int_{zw}\Big(\Delta^{\mathrm{F}(0)}_{xz}\,i\Pi_{zw}\,\Delta_{wy}^{\mathrm{F}(0)}\:\mp\:\Delta^{\mathrm{D}(0)}_{xz}\,i\Pi^*_{zw}\,\Delta_{wy}^{\mathrm{D}(0)}\Big)\;,
\end{equation}
where
\begin{equation}
i\Pi_{zw}\ =\ -\,\frac{g^2}{2!}\big(\Delta^{\mathrm{F}(0)}_{zw}\big)^2\;,\qquad -\,i\Pi^*_{zw}\ =\ -\,\frac{g^2}{2!}\big(\Delta^{\mathrm{D}(0)}_{zw}\big)^2
\end{equation}
are the standard Feynman self-energy and its complex conjugate, respectively.

Working in momentum space, we may show that
\begin{subequations}
\label{correct}
\begin{align}
\label{deltaP}
\Deltat^{+(1)}_{p^2}\ &= \ -\,2\,\Big[\big(\Delta^{\mathcal{P}}_{p^2}\big)^2\,\Im\,\Pi_{p^2}\:-\:\pi\delta'(p^2-m^2)\,\Re\,\Pi_{p^2}{}\Big]\;,\\
\Deltat^{-(1)}_{p^2} \ &=\ 2i\Big[\big(\Delta^{\mathcal{P}}_{p^2}\big)^2\,\Re\,\Pi_{p^2}\:+\:\pi \delta'(p^2-m^2)\,\Im\,\Pi_{p^2}\Big]\;,
\end{align}
\end{subequations}
where $\Re$ and $\Im$ indicate the dispersive and absorptive parts of the Feynman self-energy, and we have used the identity
\begin{equation}
\big(\Delta^{\mathrm{F}}_p\big)^2\ =\ \big(\Delta^{\mathrm{D}}_p\big)^{2*}\ =\ \big(\Delta^{\mathcal{P}}_p\big)^2\:-\:i\pi\delta'(p^2-m^2)\;,
\end{equation}
with $\delta'(p^2-m^2)\:=\:\partial_{p_0^2}\delta(p^2-m^2)$, satisfying $\int\!\D x\, f(x)\delta'(x) = -\int\!\D x\, f'(x)\delta(x)$.

Note that the one-loop correction to $\Deltat^+_{p^2}$ in~\eqref{deltaP} depends on the derivative of the dispersive part of the self-energy $\partial_{p_0^2}\,\Re\,\Pi_{p^2}$. Introducing a UV cut-off $\Lambda$, the logarithmic divergence
\begin{equation}
\frac{g^2}{32\pi^2}\ln\frac{\Lambda^2}{m^2}\ \subset\ \Re\,\Pi_{p^2}
\end{equation}
is constant with respect to $p_0^2$. As a result, the one-loop correction $\Deltat^{+(1)}_{p^2}$ is UV finite. We will discuss this suppression of the leading UV divergences in the next section.

Finally, we may calculate the resummed $\Delta^{+}_{p^2}$ and $\Delta^{-}_{p^2}$ propagators, cf.~\eqref{FDprops},
\begin{subequations}
\begin{align}
\label{resum+}
\Delta^{+}_{p^2}\ &=\  \Delta^{\mathrm{F}(0)}_{p^2}\sum_{n\,=\,0}^{\infty}\big(i\Pi_{p^2}\Delta^{\mathrm{F}(0)}_{p^2}\big)^n\ =\  \frac{i}{p^2-m^2+\Pi_{p^2}}\;,\\
\label{resum-}
\Delta^{-}_{p^2}\ &=\ -\,\Delta^{\mathrm{D}(0)}_{p^2}\sum_{n\,=\,0}^{\infty}\big(i\Pi^*_{p^2}\Delta^{\mathrm{D}(0)}_{p^2}\big)^n\ =\ \frac{i}{p^2-m^2-\Pi_{p^2}^*}\;.
\end{align}
\end{subequations}
Note that $\Delta^{-}_{p^2}$ is not the negative of the resummed Dyson propagator $\Delta^{\mathrm{D}}_{p^2} = -\,i/(p^2-m^2+\Pi_{p^2}^*)$, due to the alternating sign between each self-energy insertion. Finally, the dressed $\Deltat^{+}_{p^2}$ and $\Deltat^{-}_{p^2}$ propagators take the forms
\begin{subequations}
\begin{align}
\Deltat^{+}_{p^2}\ &=\ \Delta^+_{p^2}\:+\:\Delta^-_{p^2}\ =\ \frac{2i(p^2-m^2+i\,\Im\,\Pi_{p^2})}{(p^2-m^2+i\,\Im\,\Pi_{p^2})^2-(\Re\,\Pi_{p^2})^2}\;,\\
\Deltat^{-}_{p^2}\ &=\ \Delta^+_{p^2}\:-\:\Delta^-_{p^2}\ =\ \frac{-2i\,\Re\,\Pi_{p^2}}{(p^2-m^2+i\,\Im\,\Pi_{p^2})^2-(\Re\,\Pi_{p^2})^2}\;.
\end{align}
\end{subequations}

\section{UV sensitivity}
\label{effact}

In order to investigate further the behaviour of the leading UV divergences, we consider the one-particle irreducible (1PI) effective action~\cite{Jackiw:1974cv}
\begin{equation}
\varGamma[\varphi^{\pm}]\ =\ S[\varphi^{\pm}]\:+\:\varGamma_{1}[\varphi^{\pm}]\;,
\end{equation}
where $\varphi^{\pm}_x$ are background fields, satisfying the classical equations of motion
\begin{equation}
\label{cl}
\big(\Box_x\:+\:m^2\big)\varphi^{\pm}_x\:+\:\frac{g}{2}\,(\varphi_x^{\pm})^2\ =\ J_x\;.
\end{equation}
In terms of the renormalized tilde fields $\widetilde{\varphi}_{R,x}^{\pm} = Z_{\phi}^{-1/2}(\varphi_x^++\varphi_x^-)$, we have the equations of motion
\begin{equation}
\big(\Box_x\:+\:\overline{m}^2\big)\widetilde{\varphi}^{+}_{R,x}\:+\:\frac{\bar{g}}{2}\,(\widetilde{\varphi}_{R,x}^{+})^2\ =\ \bar{J}_x\;,\qquad
\big(\Box_x\:+\:\overline{m}^2\big)\widetilde{\varphi}^{-}_{R,x}\:+\:\frac{\bar{g}}{2}\,(\widetilde{\varphi}_{R,x}^{-})^2\ =\ 0\;.
\end{equation}

The leading quantum corrections are obtained from the functional determinant
\begin{equation}
\varGamma_1[\varphi^{\pm}]\ =\ \frac{i}{2}\:\mathrm{tr}_x\,\ln\,\mathrm{det}_{ab}\,G_{ac}^{-1}[\varphi^{\pm};x]\,G_{cb}[0;x]\;,
\end{equation}
where $G_{11(22)}^{-1}[\varphi^{\pm};x] = -\,\Box_x-m^2+(-)i\epsilon-g\varphi_x^{+(-)}$. For constant background fields $\varphi^{\pm}_x=\varphi^{\pm}$, we have
\begin{equation}
\varGamma_1[\varphi^{\pm}]\ =\ \frac{i\Omega}{2}\sum_{a\,=\,\pm}\int_k \ln \bigg[1-\frac{g\,\varphi^{a}}{k^2-m^2+a i\epsilon}\bigg]\;,
\end{equation}
where $\Omega$ is the space-time four-volume. The effective potential $V^{\mathrm{eff}}_{\varphi^{\pm}}=-\,\Omega^{-1}\varGamma[\varphi^{\pm}]$ then contains the following terms quadratic in the background field $\varphi^{\pm}$:
\begin{equation}
\label{effpot1}
V^{\mathrm{eff}}_{\varphi^{\pm}}\ \supset\ \frac{1}{2}\sum_{a\,=\,\pm}\bigg[m^2\:+\:i\,\frac{g^2}{2}\!\int_k\bigg(\frac{1}{k^2-m^2+a i\epsilon}\bigg)^{\!2}\:\bigg](\varphi^a)^2\;.
\end{equation}
Since the two vacuum graphs in~\eqref{effpot1} contain opposite pole prescriptions, they acquire a relative sign on Wick rotation. Hence, in terms of the renormalized tilde fields $\phit^{\pm}_R$, the quadratic term in the 1PI effective action is given by
\begin{align}
V^{\mathrm{eff}}_{\varphi^{\pm}}\ \supset\ \frac{1}{2}\bigg[\overline{m}^2\big[(\widetilde{\varphi}_R^+)^2
+(\widetilde{\varphi}^-_R)^2\big]\:+\:2i\bar{g}^2\,\widetilde{\varphi}_R^+\,\widetilde{\varphi}_R^-\int_k\bigg(\frac{1}{k^2-\overline{m}^2+i\epsilon}\bigg)^{\!2}\:\bigg]\;.
\end{align}
The effective masses of the $\widetilde{\varphi}^{\pm}_R$ are therefore
\begin{equation}
\partial^2_{\widetilde{\varphi}^{\pm}_R}V^{\mathrm{eff}}_{\varphi_R^{\pm}}\ =\ \overline{m}^2\;,
\end{equation}
in which the leading UV divergences have cancelled.

Suppose we also have a quartic self-interaction
\begin{equation}
\widehat{\mathcal{H}}^{\mathrm{int}}_x\ \supset \ \frac{\lambda}{4!}\big[(\phi_x^+)^4\:+\:(\phi_x^-)^4\big]\;.
\end{equation}
The additional contributions to the effective potential, quadratic in  $\varphi^{\pm}$, are
\begin{align}
V^{\mathrm{eff}}_{\varphi^{\pm}}\ \supset\ \frac{1}{2}\sum_{a\,=\,\pm}\bigg[i\,\frac{(-i\lambda)}{2}\!\int_k\frac{i}{k^2-m^2+ia\epsilon}\bigg](\varphi^a)^2\;.
\end{align}
In terms of $\widetilde{\varphi}_R^{\pm}$, the leading UV divergences again cancel in the quadratic terms.

We do not anticipate that this cancellation of vacuum graphs will persist when the propagators are dressed, as is the case in the CJT effective action~\cite{Cornwall:1974vz}. However, any UV sensitivity will be suppressed by additional factors of the coupling, modifying the renormalization-group running.

Finally, since the leading UV divergences arise in the dispersive part of the self-energy, a naive extension of this construction to finite temperature would lead to the cancellation also of the expected thermal mass. Instead, if we suppose that the canonical ensemble comprises only $+$ type states, then only the $+$ type time-ordered propagator will obtain a thermal correction, taking the form
\begin{equation}
\Delta^{+(0)}_{p^2}\ =\ \frac{i}{p^2-m^2+i\epsilon}\:+\:2\pi\delta(p^2-m^2)f_{\beta}(E_{\mathbf{p}})\;,
\end{equation}
where $f_{\beta}(E_{\mathbf{p}})=(e^{\beta E_{\mathbf{p}}}+1)^{-1}$ is the Bose-Einstein distribution. Calculating the tadpole self-energy in the high-temperature limit, $T\gg m$, gives
\begin{equation}
\partial^2_{\widetilde{\varphi}^{\pm}_R}V^{\mathrm{eff}}_{\varphi_R^{\pm}}\ =\ \overline{m}^2\:+\:\frac{\bar{\lambda} T^2}{24}\;,
\end{equation}
in which the non-trivial tree-level renormalization has now conspired to give the correct thermal mass. Here, we have used the renormalization condition $\partial^4 V(\phi^{\pm})/\partial (\phit^+_x)^4=\lambda/Z_{\lambda}$, giving $Z_{\lambda}=8$ and $Z_{\phi}\lambda = Z_{\lambda}Z_{\phi}^{-1}\bar{\lambda}=4\bar{\lambda}$. Since the Bose-Einstein distribution is tempered in the UV, no additional divergences are encountered. However, having broken the democracy of $+$ and $-$ contributions, the thermal parts of coordinate-space tree-level source-to-source transition amplitudes will not vanish for space-like separations. Nevertheless, this is as we would expect for an equilibrium ensemble, which should be space-like correlated.

\section{Electroweak oblique corrections}
\label{sec:EW}
Now we turn our attention to the loop corrections to the Standard Model, as embodied in the oblique corrections to the electroweak gauge-boson propagators~\cite{Kennedy:1988sn,Peskin:1991sw}. Despite the fact that the strictly one-loop amplitudes of the negative-energy theory are purely real, we will find agreement with the Standard Model as a result of the breakdown of naive perturbation theory due to resonance effects.

The matrix elements of the neutral- ($\NC$) and charged-current ($\CC$) interactions of the Standard Model can be written as~\cite{Peskin:1991sw}
\begin{subequations}
\label{reacs}
\begin{align}
\mathcal{M}_{\NC}\ &= \ e^2QG_{AA}Q'\:+\:\frac{e^2}{s^2c^2}\,\big(I_3-s^2Q\big)G_{ZZ}(I_3'-s^2Q')\nonumber\\& \qquad +\:\frac{e^2}{2s^2}\,\big[Q\big(I_3'-s^2Q'\big)+\big(I_3-s^2Q\big)Q'\big]G_{ZA}\;,\\
\mathcal{M}_{\CC}\ &=\  \frac{e^2}{2s^2}\,I_+I_-G_{WW}\;,
\end{align}
\end{subequations}
where $e$ is the electron charge, $s\equiv\sin\theta_W$ \& $c\equiv\cos\theta_W$ ($\theta_W$ is the Weinberg angle), $(I_3,Q)$  \& $(I_3',Q')$ are the $SU(2)$ and electric charges of the external fermions, and $I_{\pm}$ are the isospin ladder operators. The resummed propagators are given by~\cite{Kennedy:1988sn,Peskin:1991sw}
\begin{gather}
G_{A}\ =\ \frac{1}{q^2+\Pi_{A}}\:+\:\frac{\Big(\frac{\Pi_{ZA}}{q^2+\Pi_{A}}\Big)^2}{q^2-m_Z^2+\Pi_{Z}+\frac{\Pi_{ZA}^2}{q^2+\Pi_{A}}}\;,\qquad 
G_{ZA}\ =\ -\:\frac{\frac{\Pi_{ZA}}{q^2+\Pi_{A}}}{q^2-m_Z^2+\Pi_{Z}+\frac{\Pi_{ZA}^2}{q^2+\Pi_{A}}}\;,\nonumber\\
G_{Z}\ =\ \frac{1}{q^2-m_Z^2+\Pi_{Z}+\frac{\Pi_{ZA}^2}{q^2+\Pi_{A}}}\;,\qquad 
G_{W}\ =\ \frac{1}{q^2-m_W^2+\Pi_{W}}\;,
\label{Gs}
\end{gather}
where we truncate repeated subscripts for notational convenience when no ambiguity results, i.e.~$G_{A}\equiv G_{AA}$. Inserting~\eqref{Gs} into~\eqref{reacs}, the standard NC and CC matrix elements are
\begin{equation}
\mathcal{M}_{\NC}\ =\  \frac{e^2QQ'}{q^2+\Pi_{A}}\:+\:\frac{e^2}{s^2c^2}\,\frac{\big(I_3-s_*^2Q\big)\big(I_3'-s_*^2Q'\big)}{q^2-m_Z^2+\Pi_Z+\frac{\Pi_{Z\! A}^2}{q^2+\Pi_A}}\;,\qquad
\mathcal{M}_{\CC}\ =\  \frac{e^2}{2s^2}\,\frac{I_+I_-}{q^2-m_W^2+\Pi_{W}}\;,
\end{equation}
where
\begin{equation}
s_*^2\ \equiv\ s^2\:+\:sc\,\frac{\Pi_{ZA}}{q^2+\Pi_{A}}\;.
\end{equation}

The $+$ type propagators are given directly by~\eqref{Gs} and the $-$ type propagators are obtained from~\eqref{Gs} by the replacement $\Pi \to -\,\Pi^*$, cf.~\eqref{resum+} and~\eqref{resum-}. Thus, combining the $+$ and $-$ contributions, the $\CC$ matrix element instead takes the form
\begin{equation}
\mathcal{M}_{\CC} \ =\  \frac{e^2}{2s^2}\,I_+I_-\,\frac{2\big(q^2-m_W^2+i\,\Im\,\Pi_{W}\big)}{\big(q^2-m_W^2+i\,\Im\,\Pi_{W}\big)^2-\big(\Re\,\Pi_{W}\big)^2}\;.
\end{equation}
The pole mass $\overline{m}_W$ is defined via
\begin{equation}
\label{Wgap}
\big[\big(q^2-m_W^2+i\,\Im\,\Pi_{W}\big)^2\:-\:\big(\Re\,\Pi_{W}\big)^2\big]_{q^2\,=\,\overline{m}_W^2}\ = \ 0\;,
\end{equation}
i.e. $\overline{m}_W^2 = m_W^2-\Pi_{W}(\overline{m}_W^2)$, 
where it is necessary to choose the negative root of $(\Re\,\Pi_{W})^2$.
Expanding about $\overline{m}_W$, we have
\begin{equation}
\mathcal{M}_{\CC}\ =\ \frac{e^2}{2s^2}\,I_+I_-\,\frac{Z_W}{q^2-\overline{m}_{W*}^2}\;,
\end{equation}
where the running pole mass $\overline{m}_{W*}(q^2)$ is defined such that $\frac{\D}{\D q^2}\,\overline{m}_{W*}^2(q^2)\big|_{q^2\,=\,\overline{m}_W^2}\ =\ 0$ and $\overline{m}_{W*}^2(\overline{m}_W^2)\ =\ \overline{m}_W^2$. The wavefunction renormalization $Z_W$ is given by
\begin{equation}
\label{ZW}
Z_W^{-1}\ =\ \frac{\D}{\D q^2}\,\frac{\big(q^2-m_W^2+i\,\Im\,\Pi_{W}\big)^2-\big(\Re\,\Pi_{W}\big)^2}{2\big(q^2-m_W^2+i\,\Im\,\Pi_{W}\big)}\,\Bigg|_{q^2\,=\,\overline{m}_{W}^2}\;,
\end{equation}
and this appears to differ significantly from the standard expression:
\begin{equation}
Z_W^{-1}\ =\ \frac{\D}{\D q^2}\,\big(q^2-m^2+\Pi_W\big)\Big|_{q^2\,=\,\overline{m}_{W}^2}\;.
\end{equation}
Nevertheless, performing the $q^2$ derivative in~\eqref{ZW} and using the gap equation~\eqref{Wgap}, we do in fact find
\begin{equation}
\label{ZWfinal}
Z_W^{-1}\ =\ 1\:+\:\frac{\D \Pi_W}{\D q^2}\,\bigg|_{q^2\,=\,\overline{m}_W^2}~,
\end{equation}
which is in agreement with the standard result. Importantly, the full
dispersive correction from $\mathrm{Re}\,\Pi_W$ is present
in~\eqref{ZWfinal}. This should be compared with the explicit one-loop
result in~\eqref{deltaP}. Since $\mathrm{Re}\,\Pi_W$ appears only
quadratically in~\eqref{ZW}, the one-loop dispersive correction has
arisen as a result of being near resonance,
i.e.~$\mathrm{Re}\,\Pi_{W}\sim -\,(q^2-m^2_W+i\mathrm{Im}\,\Pi_W)$,
thereby reducing the naive perturbative order.  Similar resonance
effects give rise to the double-counting problem in semi-classical
transport phenomena~\cite{Kolb:1980}. Note that the limit $\Pi\to 0$
is somewhat delicate, giving rise to $Z_{W}= 2$ at leading order (see Section~\ref{sec:matrix}).

For the pure photon contribution to the $\NC$ matrix element, we have
\begin{equation}
\mathcal{M}_{\NC}\ \supset\ e^2QQ'\,\frac{2(q^2+i\,\Im\,\Pi_A)}{\big(q^2+i\,\Im\,\Pi_A\big)^2-\big(\Re\,\Pi_A)^2}\;.
\end{equation}
Again, we expand the denominator around the pole, this time at $q^2=0$. Since $\Pi_A=0$ at $q^2=0$ by gauge invariance, see~\cite{Peskin:1991sw}, we find
\begin{equation}
\mathcal{M}_{\NC}\ \supset\ e_*^2Q\,\frac{1}{q^2}\,Q'\;,\qquad 
e_*^2\ \equiv\ e^2\bigg[1\:+\:\frac{\D \Pi_A}{\D q^2}\,\bigg|_{q^2\,=\,0}\:\bigg]^{-1}\;,
\end{equation}
also in agreement with the standard result.

For the $Z$-boson contribution, we neglect the terms quadratic in $\Pi_{ZA}$ in the denominators of~\eqref{Gs}~\cite{Peskin:1991sw} and, proceeding as above, find 
\begin{equation}
Z_Z^{-1} \ = \ \frac{\D}{\D q^2}\,\frac{\big(q^2-m_Z^2+i\,\Im\,\Pi_Z\big)^2-\big(\Re\,\Pi_Z\big)^2}{2\big(q^2-m_Z^2+i\,\Im\,\Pi_Z\big)}\,\Bigg|_{q^2\,=\,\overline{m}_Z^2}\;.
\end{equation}
Since this can also be written as
\begin{equation}
Z_Z^{-1}\ =\ \frac{\D}{\D q^2}\,\frac{\big(q^2-m_Z^2+i\,\Im\,\Pi_Z\big)^2-\big(\Re\,\Pi_Z\big)^2}{2\big(\!-\Re\,\Pi_Z\big)}\,\Bigg|_{q^2\,=\,\overline{m}_Z^2}\;,
\end{equation}
it follows that the sum of terms involving the Z boson gives
\begin{equation}
\label{ZNC}
\mathcal{M}_{\NC}\ \supset\ \frac{e^2}{s^2c^2}\,\big(I_3-s_*^2Q\big)\,\frac{Z_Z}{q^2-\overline{m}_{Z*}^2}\,\big(I_3'-s_*^2Q'\big)\;,\qquad Z_Z^{-1}\ =\ 1\:+\:\frac{\D \Pi_Z}{\D q^2}\,\bigg|_{q^2\,=\,\overline{m}_Z^2}\;,
\end{equation}
again with the usual wavefunction renormalization.

Finally, defining the running wavefunction renormalizations $Z_{Z*}(q^2)$ and $Z_{W*}(q^2)$ in the usual way, i.e.
\begin{equation}
\frac{e_*^2}{s_*^2c_*^2}\,Z_{Z*}\ \equiv \ \frac{e^2}{s^2c^2}\,Z_{Z}\;,\qquad \frac{e_*^2}{s_*^2}\,Z_{W*}\ \equiv \ \frac{e^2}{s^2}\,Z_{W}\;,
\end{equation}
with $c_*^2=1-s_*^2$, the complete $\NC$ and $\CC$ matrix elements take their well-known forms~\cite{Peskin:1991sw}
\begin{equation}
\mathcal{M}_{\NC}\ =\ Q\frac{e_*^2}{q^2}Q'\:+\:\frac{e_*^2}{s_*^2c_*^2}\,\frac{\big(I_3-s_*^2Q\big)Z_{Z*}\big(I_3'-s_{*}^2Q'\big)}{q^2-\overline{m}_{Z*}^2}\;,\qquad 
\mathcal{M}_{\CC}\ =\  \frac{e_*^2}{2s_*^2}\,I_+\,\frac{Z_{W*}}{q^2-\overline{m}_{W*}^2}\,I_-\;.
\end{equation}

\section{Conclusions}

We have described a departure from standard quantum field theory, which permits both positive- and negative-energy states. At tree-level, the theory gives rise to manifestly-causal source-to-source amplitudes that are in agreement with the standard S-matrix theory. It also gives rise to the cancellation of leading UV divergences and the absence of a zero-point energy. However, the theory appears to deviate significantly from standard quantum field theory at the loop level, casting doubts on its usefulness. Even so, there is a quite remarkable agreement with the one-loop electroweak corrections to the Standard Model that is a consequence of a breakdown of naive perturbation theory.

\ack

This work was supported in part by the Lancaster-Manchester-Sheffield Consortium for Fundamental Physics under STFC grant ST/J000418/1 and the IPPP under STFC grant ST/G000905/1. The work of P.M.~is supported by a University Foundation Fellowship (TUFF) from the Technische Universit\"{a}t M\"{u}nchen and by the Deutsche Forschungsgemeinschaft (DFG) cluster of excellence Origin and Structure of the Universe. P.M.~would like to thank High Energy Physics at Imperial College London, Theoretical Physics at the University of Liverpool, and the organizers and delegates of DISCRETE2014 for their hospitality, and helpful questions, comments and suggestions.


\section*{References}

\begin{thebibliography}{99}

\bibitem{Fermi}
  E.~Fermi, Rev.~Mod.~Phys.~4 (1932) 87.

\bibitem{Shirokov}
M.~I.~Shirokov, Yad.~Fiz.~4 (1966) 1077 [Sov.~J.~Nucl.~Phys.~4
(1967) 774].

\bibitem{Hegerfeldt:1993qe}
  G.~C.~Hegerfeldt,
  Phys.\ Rev.\ Lett.\  {\bf 72} (1994) 596.

\bibitem{Kennedy:1988sn}
  D.~C.~Kennedy and B.~W.~Lynn,
  Nucl.\ Phys.\ B {\bf 322} (1989) 1.
  
\bibitem{Peskin:1991sw}
  M.~E.~Peskin and T.~Takeuchi,
  Phys.\ Rev.\ D {\bf 46} (1992) 381.
 
\bibitem{Elze:2005kt}
  H.~T.~Elze,
  Int.\ J.\ Theor.\ Phys.\  {\bf 46} (2007) 2063
  [hep-th/0510267].
  
\bibitem{Takahasi:1974zn}
  Y.~Takahasi and H.~Umezawa,
  Collect.\ Phenom.\  {\bf 2} (1975) 55.

\bibitem{Arimitsu:1985xm}
  T.~Arimitsu and H.~Umezawa,
  Prog.\ Theor.\ Phys.\  {\bf 77} (1987) 32.
  
\bibitem{bicomplex}
        D.~Alpay,
        M.~E.~Luna-Elizarrar\'{a}s,
        M.~Shapiro and
        D.~C.~Struppa, \emph{Basics of functional analysis with bicomplex scalars, and bicomplex schur analysis}, Springer Briefs in Mathematics, Springer (2014).
        
\bibitem{Bender:2005tb}
  C.~M.~Bender,
  Contemp.\ Phys.\  {\bf 46} (2005) 277
  [quant-ph/0501052].
  
\bibitem{Millington:2012pf}
  P.~Millington and A.~Pilaftsis,
  Phys.\ Rev.\ D {\bf 88} (2013) 085009 [arXiv:1211.3152 [hep-ph]].
  
  \bibitem{Dickinson:2013lsa}
  R.~Dickinson, J.~Forshaw, P.~Millington and B.~Cox,
  JHEP 06 (2014) 049 [arXiv:1312.3871 [hep-th]]; R.~Dickinson, J.~Forshaw and P.~Millington (2015), \emph{in preparation}.
  
\bibitem{Bloch:1937pw}
  F.~Bloch and A.~Nordsieck,
  Phys.\ Rev.\  {\bf 52} (1937) 54.
  
\bibitem{Lehmann:1954rq}
  H.~Lehmann, K.~Symanzik and W.~Zimmermann,
  Nuovo Cim.\  {\bf 1} (1955) 205.
  
\bibitem{Schwinger:1960qe}
      J.~S.~Schwinger,
      J.~Math.~Phys.~2 (1961) 407--432.
      
\bibitem{Keldysh:1964ud}
      L.~V.~Keldysh,
      Zh.~Eksp.~Teor.~Fiz.~47 (1964) 1515--1527.  

\bibitem{Berges:2004yj}
  J.~Berges,
  AIP Conf.\ Proc.\  {\bf 739} (2005) 3 [hep-ph/0409233].
  
\bibitem{Jackiw:1974cv}
  R.~Jackiw,
  Phys.\ Rev.\ D {\bf 9} (1974) 1686.

\bibitem{Cornwall:1974vz}
  J.~M.~Cornwall, R.~Jackiw and E.~Tomboulis,
  Phys.\ Rev.\ D {\bf 10} (1974) 2428.

\bibitem{Kolb:1980} 
  E.~W.~Kolb and S.~Wolfram,
  Nucl.\ Phys.\ B {\bf 172}, 224 (1980)
  [Erratum-ibid.\ B {\bf 195}, 542 (1982)].

\end{thebibliography}
\end{document}